\documentclass[aps,showpacs,superscriptaddress,twocolumn]{revtex4}%
\usepackage{amsfonts}
\usepackage{amsmath}
\usepackage{amssymb}
\usepackage{graphicx}%
\begin{document}

\title{Molecular dynamics study of hcp/fcc nucleation and growth in bcc iron driven by uniaxial
strain}
\author{Baotian Wang}
\affiliation{Institute of Theoretical Physics and Department of
Physics, Shanxi University, Taiyuan 030006, People's Republic of
China} \affiliation{Institute of Applied Physics and Computational
Mathematics, P.O. Box 8009, Beijing 100088, People's Republic of
China}
\author{Jianli Shao}
\affiliation{Institute of Applied Physics and Computational
Mathematics, P.O. Box 8009, Beijing 100088, People's Republic of
China}
\author{Guangcai Zhang}
\affiliation{Institute of Applied Physics and Computational
Mathematics, P.O. Box 8009, Beijing 100088, People's Republic of
China}
\author{Weidong Li}
\affiliation{Institute of Theoretical Physics and Department of
Physics, Shanxi University, Taiyuan 030006, People's Republic of
China}
\author{Ping Zhang}
\thanks{To whom correspondence should be
addressed. E-mail: zhang\_ping@iapcm.ac.cn} \affiliation{Institute
of Applied Physics and Computational Mathematics, P.O. Box 8009,
Beijing 100088, People's Republic of China} \affiliation{Center for
Applied Physics and Technology, Peking University, Beijing 100871,
People's Republic of China}
\date{\today}


\begin{abstract}
Molecular dynamics simulations are performed to investigate the
structural phase transition in body-centered cubic (bcc) single
crystal iron under high strain rate loading. We study the nucleation
and growth of the hexagonal-close-packed (hcp) and
face-centered-cubic (fcc) phases, and their crystal orientation
dependence. Results reveal that the transition pressures are less
dependent on the crystal orientations ($\mathtt{\sim}$14 GPa for
loading along [001], [011], and [111] directions). However, the
pressure interval of mixed phase for [011] loading is much shorter
than loading along other orientations. And the temperature increased
amplitude for [001] loading is evidently lower than other
orientations. The hcp/fcc nucleation process is presented by the
topological medium-range-order analysis. For loading along [001]
direction, we find that the hcp structure occurs firstly and grows
into laminar morphology in the (011)$_{\text{bcc}}$ planes with a
little fcc atoms as intermediate structure. For loading along [011]
and [111] directions, both the hcp and fcc structures nucleation and
growth along the \{110\}$_{\text{bcc}}$ planes are observed, whose
morphology is also discussed.

\end{abstract}
\maketitle

\section{INTRODUCTION}

Besides melting, dislocation generation, and twinning, structural
phase transition is also one important physical process in
understanding the plasticity of solid (metallic, particularly)
materials. Among various metallic materials, iron, due to its
critical importance in condensed-matter physics, materials science,
geophysics, and human development, have been extensively studied in
aiming at understanding its structures and phase transition under
extreme conditions. Thereinto, the research on classical phase
transition of iron from ferromagnetic body-centered cubic (bcc)
ground state $\alpha$ phase to nonmagnetic hexagonal-close-packed
(hcp) $\epsilon$ phase has a history of more than half a century.
Early representative work was carried out in 1956 by Bancroft
\emph{et al.} \cite{Bancroft}, where they discovered the transition
under shock wave compression at pressure around 13 GPa. Thereafter,
plenty of theoretical and experimental works have been done to study
the structure properties, inner physics, and mechanical features
involving this $\alpha\mathtt{\rightarrow}\epsilon$ transition
process
\cite{Jamieson1962,Barker1974,Taylor,Boettger,Hasegawa,WangFM,Caspersen}.
Among these studies, in 1962, Jamieson and Lawson obtained
crystallographic evidence for the
$\alpha\mathtt{\rightarrow}\epsilon$ transition at 13 GPa in static
x-ray diffraction experiments \cite{Jamieson1962}, which confirmed
previous shock-wave experiment of Bancroft \emph{et al.}
\cite{Bancroft}; in 1991, Taylor and Pasternak reported the onset of
the transition in range of 9-15 GPa with  a large hysteresis width
of about 6 GPa \cite{Taylor}; in 1997, Boettger and Wallace
presented theoretical analysis of metastability and dynamics of the
transition \cite{Boettger}; in 1999, Wang and Ingalls suggested
three possible models for the $\alpha\mathtt{\rightarrow}\epsilon$
transition mechanism \cite{WangFM}; and in 2004, Caspersen \emph{et
al.} theoretically investigated the influence of shear in the
transition based on a multiscale model \cite{Caspersen}.

However, the structural transition is a mutation process, which
leads to direct observation of the phase transition very difficult
in experiment. And the abrupt changes in physical and mechanical
characteristics also bring about difficulty in theoretically
analyzing the dynamic response and micro-behavior of materials.
Therefore, it is necessary to carry out systematic simulations at
atomic level by using molecular dynamics (MD). MD is a method for
computationally evaluating the thermodynamic and transport
properties of materials by solving the classical equations of motion
at the atomic level. Many preceding works have testified that MD is
the most conceptually straightforward method and an altogether very
utilitarian method for obtaining atomic details. In 2002, Kadau
\emph{et al.} \cite{Kadauscience} realized the micro-simulation of
shock-induced phase transition of solid iron along [001] direction
by employing MD method with an embedded atom method (EAM) potential
\cite{Daw1983,Daw1984}, showing the micro-features of
$\alpha\mathtt{\rightarrow}\epsilon$ transition. Subsequent
ultrafast x-ray diffraction experiments processed by Kalantar
\emph{et al.} \cite{Kalantar} verified the validity of this work. In
2005, Kadau \emph{et al.} \cite{KadauPRB} simulated shock-induced
phase transformations in bcc iron induced by shock loading along
[001], [011], and [111] directions to research the orientation
dependence of the developing microstructure on the crystallographic
shock direction. They found that the fcc phase appeared especially
when loading along [011] and [111] directions. In 2007, Kadau
\emph{et al.} \cite{KadauPRL} utilized the MD method again to
simulate the polycrystalline iron in shock loading condition,
simulation results correspond properly with experiment
\cite{Yaakobi}.

These abovementioned MD studies have greatly enriched our knowledge
of the structural transition of iron and, needless to say,
subsequently inspired more simulation works \cite{Cui,Shao}. In our
last work \cite{Shao}, we have simulated the
$\alpha\mathtt{\rightarrow}\epsilon$ structural transition of iron
under isothermal compression along the [001] direction. It was
showed that the laminar structure forms along \{110\}$_{\text{bcc}}$
planes. On the other hand, when the loading direction is different,
both the hcp and fcc nucleation and growth in bcc single crystal
iron will occur. To date, no detailed investigation has been
proceeded to study the orientation dependence of hcp and fcc
nucleation and growth. This issue is focus of our present paper.
Specially, by adiabatic compression along [001], [011], and [111]
directions, in this paper, we investigate the
$\alpha\mathtt{\rightarrow}\epsilon/\gamma$ structural phase
transition in bcc single crystal iron. The pressure and temperature
effect, nucleation mechanism and morphology characters are all
discussed.

The rest of this paper is arranged as follows. In Sec. II the
simulation method is briefly described. In Sec. III we present and
discuss our simulated results. A summary of our main results is
given in Sec. IV.

\section{METHODOLOGY OF THE SIMULATIONS}

For this work, we use classical MD and the EAM potential of iron
developed by Voter-Chen (VC) \cite{VC} to conduct the simulations.
Our calculations are performed in three cases by using three samples
A, B, and C to simulate the loadings along the [001], [011], and
[111], respectively. Specially, sample A consists of
50$_{x,[100]}$$\times$50$_{y,[010]}$$\times$100$_{z,[001]}$ atomic
cells, sample B consists of
36$_{x,[011]}$$\times$100$_{y,[\bar{1}00]}$$\times$72$_{z,[0\bar{1}1]}$
cells, and sample C consists of
36$_{x,[\bar{1}10]}$$\times$40$_{y,[\bar{1}\bar{1}2]}$$\times$180$_{z,[111]}$
cells. Periodical boundary conditions are adopted in all directions.
The original temperature in all samples is set at 60 K by speed
calibration method \cite{Hoffman}. The initial state is manipulated
at zero pressure through modulating the lattice constant. Here the
lattice constant of all samples is a$_{0}$=0.28725 nm. Loading with
high strain rate of $\mathtt{\sim}$10$^{8}$ along the [001], [011],
and [111] directions is implemented by shortening the lattice
constants of these three orientations step by step during the
nonequilibrium MD process. The Verlet algorithm \cite{Swope} is used
to integrate the equation of motion. The simulation time is up to
200 ps, with the time step set at 2 fs. Special techniques are
required to analyze the huge amount of data produced in the MD
simulations. Here, the local structure around each atom are resolved
by using the topological medium-range-order analysis
\cite{Honeycutt} technique. The phase mass fraction analysis and
radial distribution function are also presented to obtain the phase
transition information.

\section{RESULTS AND DISCUSSION}

\subsection{Pressure and temperature effects}

\begin{figure}[tbp]
\begin{center}
\includegraphics[width=0.8\linewidth]{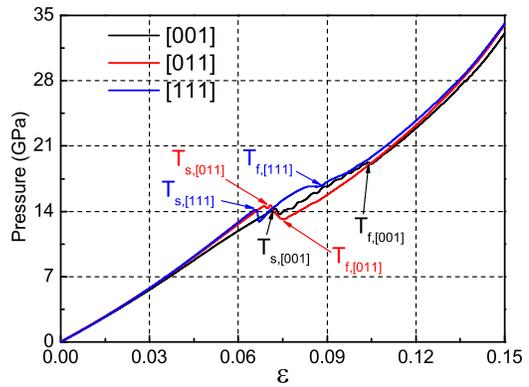}
\end{center}
\caption {(Color online) Variation of pressure with the strain
$\varepsilon$ for loading along [001], [011], and [111] directions.
The mixed phase intervals start at T$_{\text{s},[001]}$(0.072,
14.21), T$_{\text{s},[011]}$(0.071, 14.66), and
T$_{\text{s},[111]}$(0.066, 14.16) and finish at
T$_{\text{f},[001]}$(0.104, 19.25), T$_{\text{f},[011]}$(0.075,
13.17), and T$_{\text{f},[111]}$(0.088, 16.74) for [001], [011], and
[111] loadings, respectively.} \label{pressure}
\end{figure}

\begin{figure}[tbp]
\begin{center}
\includegraphics[width=0.8\linewidth]{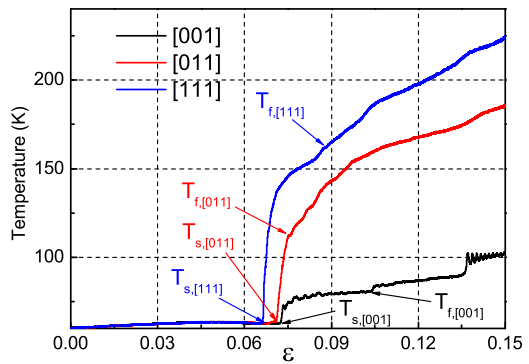}
\end{center}
\caption{(Color online) Variation of temperature with the strain
$\varepsilon$ for loading along [001], [011], and [111] directions.
The letters T$_{\text{s},i}$ and T$_{\text{f},i}$ (\emph{i} stands
for [001], [011], and [111]) have same strains with that in Fig.
\ref{pressure}.} \label{temperature}
\end{figure}

\begin{figure}[tbp]
\begin{center}
\includegraphics[width=0.8\linewidth]{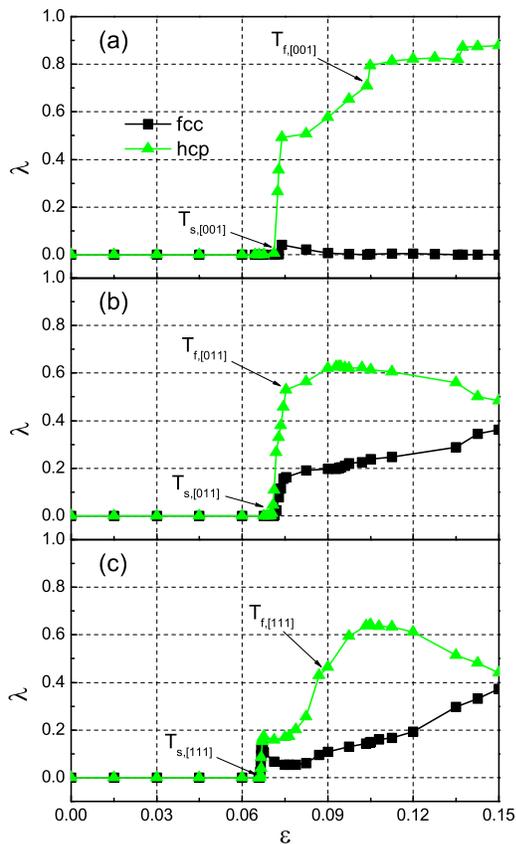}
\end{center}
\caption{(Color online) Mass fractions $\lambda$ for the fcc and hcp
phases change with strain $\varepsilon$ for loading along [001] (a),
[011] (b), and [111] (c) directions. The letters T$_{\text{s},i}$
and T$_{\text{f},i}$ (\emph{i} stands for [001], [011], and [111])
have same strains with that in Fig. \ref{pressure}.}\label{fraction}
\end{figure}

To begin with, we perform the pressure analysis during the loading
process. The calculated pressure as a function of compression strain
$\varepsilon$ is shown in Fig. \ref{pressure}, where
$\varepsilon$=1-V/V$_{0}$ and V$_{0}$ is the initial volume of the
samples. With increasing the compression strain, the following
prominent features can be seen from Fig. \ref{pressure}: (i) The
transition critical points in the strain-pressure space are
T$_{\text{s},[001]}$(0.072, 14.21), T$_{\text{s},[011]}$(0.071,
14.66), and T$_{\text{s},[111]}$(0.066, 14.16) for loading along
[001], [011], and [111] directions, respectively, which illustrates
that the critical pressures of transition are less dependent on the
crystal orientations. This is similar to the recent shock-loading
simulations \cite{KadauPRB}. However, the strain of
T$_{\text{s},[111]}$ is evidently smaller than that of other two
critical points, suggesting that sample C begins transition first.
After the critical points the pressure is lowered due to the
nucleation of the new phases; (ii) Close to the transition critical
point T$_{\text{s},[111]}$, the pressure-strain curve for [001]
loading is below other two curves. This indicates that samples B and
C experience a more evident hardening process compared with sample A
before phase transition; (iii) The mixed phase intervals are
T$_{\text{s},[001]}$ to T$_{\text{f},[001]}$, T$_{\text{s},[011]}$
to T$_{\text{f},[011]}$, and T$_{\text{s},[111]}$ to
T$_{\text{f},[111]}$ for loading along [001], [011], and [111]
directions, respectively. Obviously, the mixed phase interval in
[011] loading is much shorter compared with that in [001] and [111]
loadings. This is due to the fact that the compression direction in
sample B is parallel to the nucleation planes (see below for further
discussion).

Consistent with previous reports \cite{Hawreliak, Cui}, in our
simulations we have also observed two main steps during structural
transition of iron: firstly, atoms in nucleation planes
\{011\}$_{\text{bcc}}$ are compressed to form a hexagonal structure;
secondly, neighbor \{110\}$_{\text{bcc}}$ layers experience a
relative slip to transform into hcp structure.

Under adiabatic compression, the temperature is set free, i.e., not
controlled artificially. Thus the temperature evolution with the
strain can be calculated based on our data. Results are presented in
Fig. \ref{temperature}. One can see that the temperature values at
the three starting transition critical points (namely,
T$_{\text{s},[001]}$, T$_{\text{s},[011]}$, and
T$_{\text{s},[111]}$) are all near 64 K, which further implies that
the onset of phase transition is insensitive to the loading
orientations. Right after the onset of the phase transition,
however, loadings along the three different directions display
remarkably different temperature effects by the fact shown in Fig.
\ref{temperature} that the temperature is higher one by one with a
sequence of [001], [011], and [111] in loading orientation. While
the temperatures of samples B and C change up to 185 K and 223 K,
the temperature of sample A only increases to 102 K at the end of
loading. Also, the temperature increase style is different. The
temperature of sample A increases in an up-stair-like manner, which
can be seen more carefully near the labels
T$_{\text{s},[001]}$(0.072, 63.63), T$_{\text{f},[001]}$(0.104,
81.17), and followed by a sawtooth-like slope after exceeding the
strain of 0.136. Here, the temperature climb up the first stair may
contribute to the first step of transition mechanism. This
up-stair-like process only costs about 2 ps. Then, the temperature
keeps almost no change until the second stair. The second and third
stairs in the temperature of sample A have strong relation with the
second step in the structural transition of iron. Closer view can
find that the temperature fluctuates dramatically after the third
stair and only mildly fluctuates after the first two stairs. These
features of temperature in sample A has not been observed in samples
B and C. Instead, the temperatures in samples B and C first alter
suddenly up to 110 K and 142 K from the initial value of
$\mathtt{\sim}$64 K, respectively in 5.3 and 9.3 ps. Then they
increase steadily with strain. On the whole, the remarkable
distinction in temperature before and after phase transition
illustrates the substantial structural and energetic complexity
during transition process.

Although the main transition mechanism remains the same for the
three kinds of loadings, the details are significantly different.
This can be seen from the phase mass fraction analysis. Figures
3(a)-(c) show the mass fraction $\lambda$ for the hcp and fcc phases
during their evolution with the loading strain along the three
directions, respectively. In the overall view, the variations of the
mass fraction of close packed (fcc/hcp) phases as a function of
strain in all three samples are similar to that of temperature.
Therefore, the temperature and the nucleation of close packed phases
have tight relation. In addition, while most atoms in sample A
nucleate into the hcp structure, in samples B and C there will
emerge near half proportion of fcc atoms. This is similar to the
observation in shock-wave loading simulation \cite{KadauPRL}, where
it was found that the hcp/fcc ratio within a grain decreases the
more the shock direction deviates from the [001]$_{\text{bcc}}$
direction of the initial polycrystal. In our study, with the
pressure increasing, more fcc atoms will form when loading along
[011] and [111] directions. The hcp mass fraction decreases with the
increase of fcc phase mass fraction, which indicates that the
nucleation of fcc is due to the slip of neighboring hexagonal
layers.

\begin{figure}[tbp]
\includegraphics[width=1.0\linewidth]{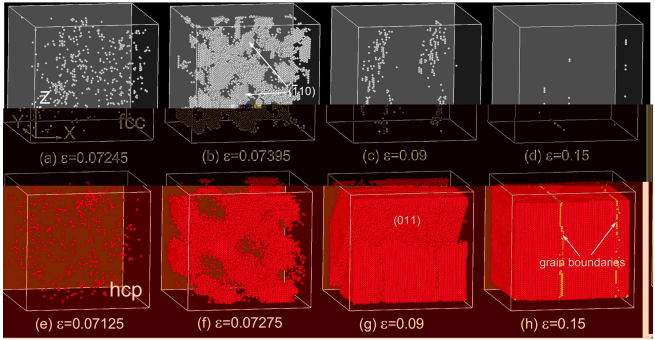}
\caption {(Color online) Morphology evolution of the fcc phase (the
first row) and the hcp phase (the second row) in sample A with
strains labeled under each snapshot. The coordinates are shown in
panel (a) and the compression directions is along \emph{Z} axis.
Color coding was obtained by using the topological
medium-range-order analysis: white: fcc, red: hcp, yellow: grain
boundaries.}\label{001}
\end{figure}

\begin{figure}[tbp]
\includegraphics[width=1.0\linewidth]{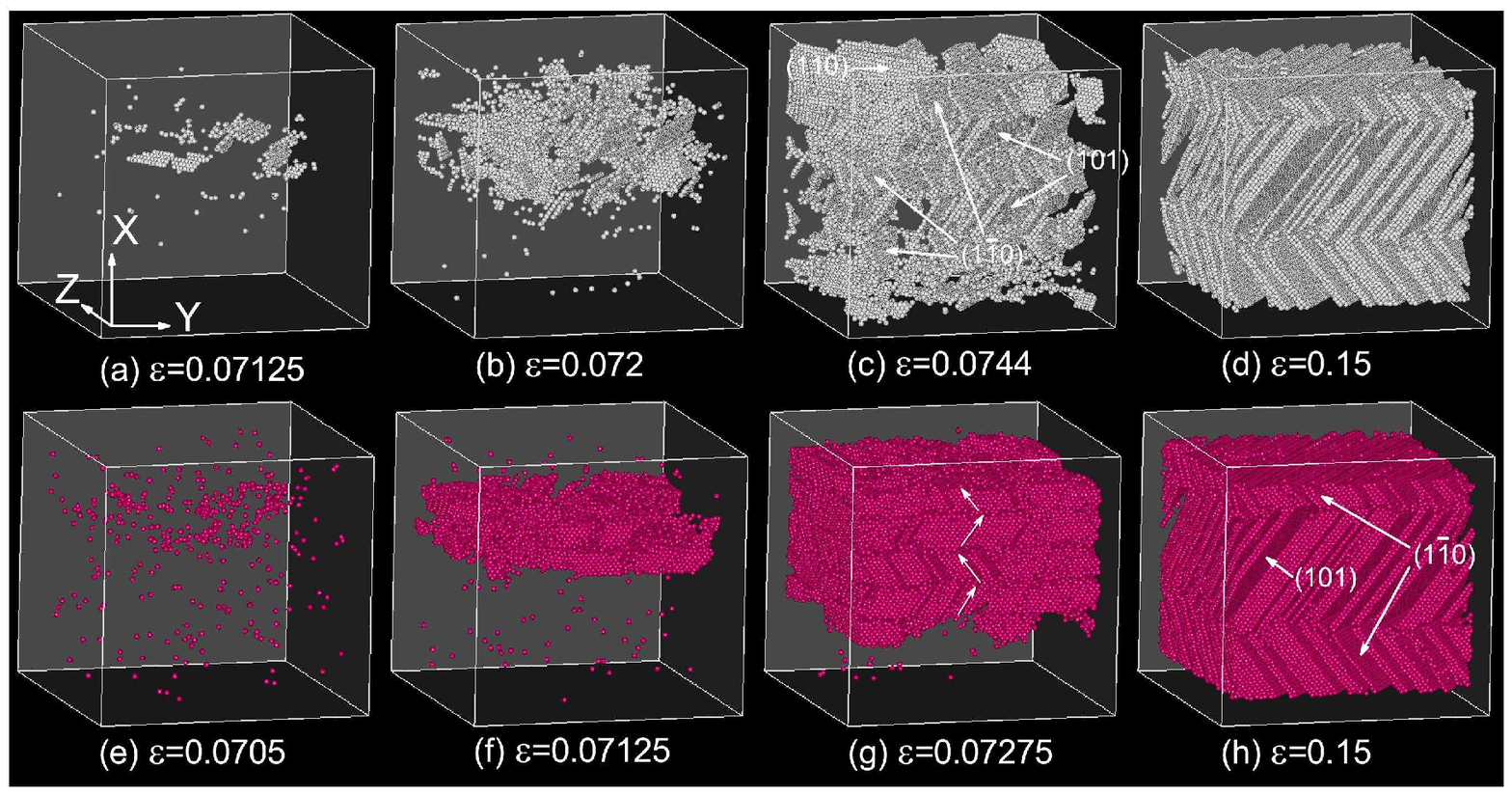}
\caption {(Color online) Morphology evolution of the fcc phase (the
first row) and the hcp phase (the second row) in sample B with
strains labeled under each snapshot. The coordinates are shown in
panel (a) and the compression directions is along \emph{X} axis.
Color criterion same with Fig. \ref{001}.}\label{011}
\end{figure}

\begin{figure}[tbp]
\begin{center}
\includegraphics[width=1.0\linewidth]{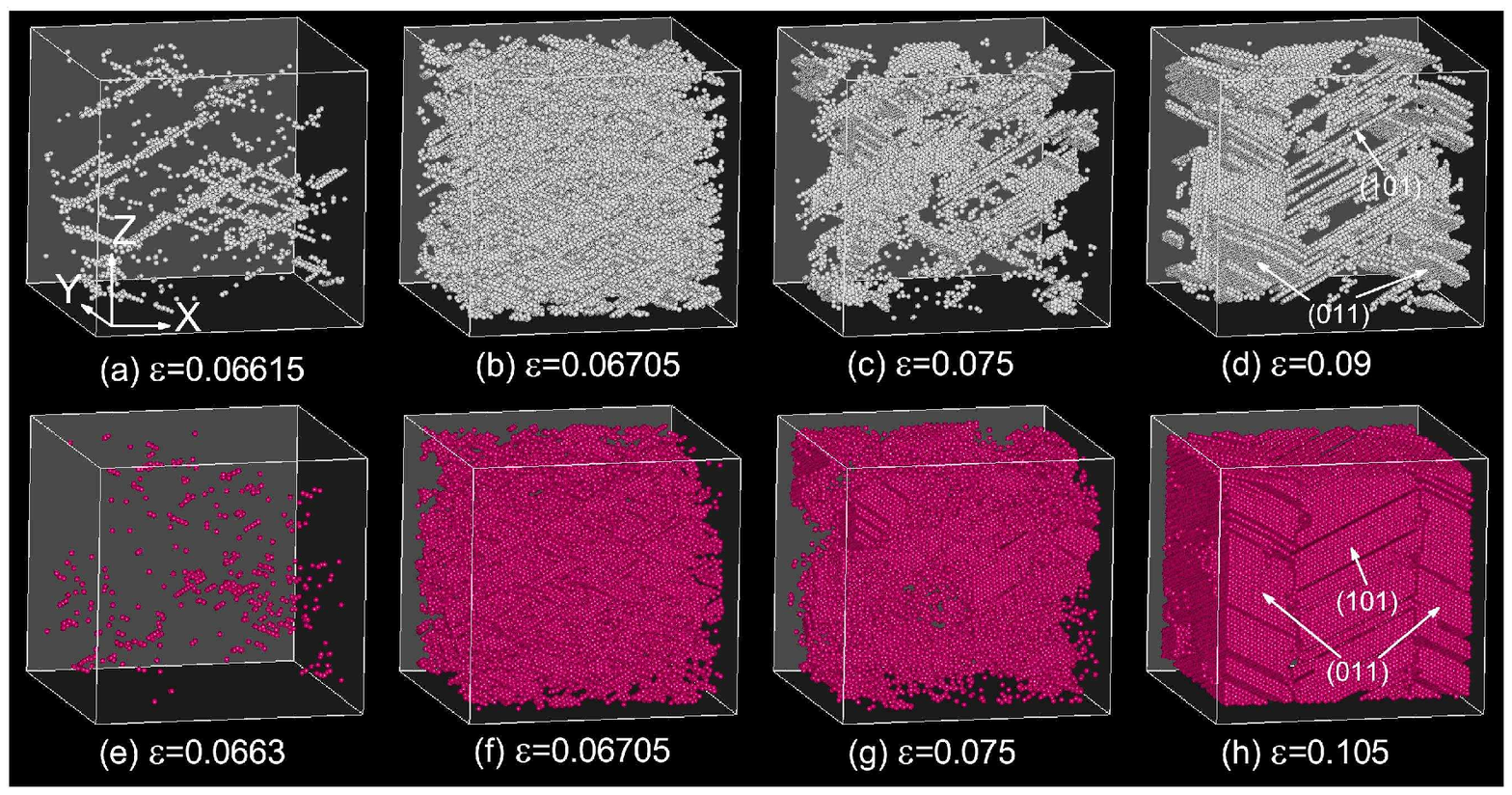}
\end{center}
\caption {(Color online) Morphology evolution of the fcc phase (the
first row) and the hcp phase (the second row) in sample C with
strains labeled under each snapshot. The coordinates are shown in
panel (a) and the compression directions is along \emph{Z} axis.
Color criterion same with Fig. \ref{001}.}\label{111}
\end{figure}

\begin{figure}[tbp]
\begin{center}
\includegraphics[width=1.0\linewidth]{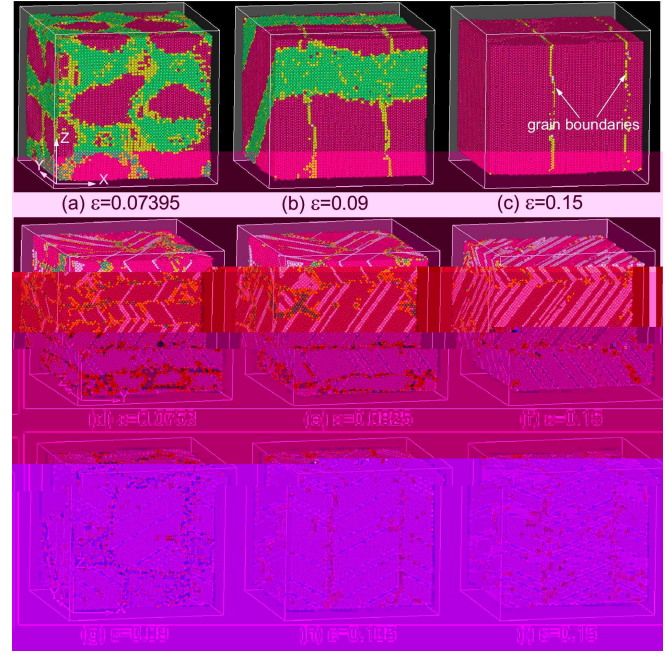}
\end{center}
\caption {(Color online) Compressed samples under different strains
with the first row (three snapshots) for [001] loading, the second
row for [011] loading, and the third row for [111] loading. The
compression directions of all samples are from top to bottom. The
coordinates are shown in (a), (d) and (g) for loading along [001],
[011], and [111] directions, respectively. The green atoms stand for
the bcc structure and other color criterion same with Fig.
\ref{001}.}\label{morphology}
\end{figure}

\subsection{Nucleation and growth of hcp and fcc phases}

The nucleation and growth of the hcp phase under [001] compression
has been investigated in Ref. \cite{Shao}. Here, we focus on the
emergence and growth of both hcp and fcc phases along different
crystal directions by continuous uniaxial compression. Over the
critical pressure, the close packed phases begin to nucleate in all
three kinds of loading. The microscopic view of the initial
nucleation and growth of the fcc (upper panels) and hcp (lower
pannels) phases for samples A, B, and C are shown in Figs. 4-6,
respectively. The evolved loading strain corresponding to these
snapshots are also labeled. For an overall view, the morphologies of
samples A, B, and C are also shown in Fig. \ref{morphology} at some
evolved strains. Based on these figures, we can clearly see that for
the [001] and [111] loadings, until strain being loaded up to 0.09,
the bcc structure atoms still maintain bulk morphology, as shown in
Fig. \ref{morphology}(b) and 7(g). For the loading along [011]
direction, on the contrary, only a few bcc atoms dispersively
distribute among bulk hcp and fcc atoms when the loaded strain is
over 0.075. This further implies that the mixed phase interval in
sample B is shorter than in other two samples.

For the [001] loading, as shown in Fig. \ref{001}, some fcc-phase
atoms only appear in mixed phase interval and nucleate on
($\bar{1}$10)$_{\text{bcc}}$ plane. With increasing pressure, the
fcc atoms will transform to hcp or grain boundary atoms. From Fig.
\ref{001}(d), we can find that a little amount of fcc atoms appear
only in the twin boundaries. For hcp atoms, initially, they
homogeneously nucleate [Fig. \ref{001}(e)], then evolve into
irregular grains [Fig. \ref{001}(f)]. These grains subsequently
develops into the typical laminar structure along the
(011)$_{\text{bcc}}$ planes, as shown in Fig. \ref{001}(g).
Ultimately, the laminar structure grows into larger and reugular
grains divided by the grain boundaries [Fig. \ref{001}(h)]. The
grain boundary planes are (100)$_{\text{bcc}}$ planes, paralleling
with the compression direction.

While loading along [011] (see Fig. \ref{011}), the fcc phase
nucleates on (110)$_{\text{bcc}}$, (101)$_{\text{bcc}}$, and
(1$\bar{1}$0)$_{\text{bcc}}$ planes and the hcp on
(101)$_{\text{bcc}}$ and (1$\bar{1}$0)$_{\text{bcc}}$ planes. These
nucleation planes all belong to the \{110\}$_{\text{bcc}}$ family.
More specially, the fcc atoms first nucleate as a flaky structure
[Fig. \ref{011}(a)], then, together with hcp atoms, evolve into the
thin samdwich structure perpendicular to the compression direction
[see Fig. \ref{011}(b), \ref{011}(c), \ref{011}(f), \ref{011}(g),
and Fig. \ref{morphology}(d)]. At the end of loading, the fcc and
hcp atoms form stacking fault structure and relatively thick
samdwich structure [Fig. \ref{011}(d) and \ref{011}(h)]. Therein,
the samdwich structure is divided by the grain boundaries, which can
be more clearly seen from the yellow atoms in Fig. \ref{morphology}.
Note that the grain boundary planes are (011)$_{\text{bcc}}$ planes
and perpendicular to the compression direction.

For loading along [111] (see Fig. \ref{111}), the nucleation and
growth of the close packed phases are different. Just over the
critical pressure, both fcc and hcp atoms nucleate to form lots of
small but long flaky structure [Fig. \ref{111}(a), \ref{111}(b),
\ref{111}(e), and \ref{111}(f)]. From Fig. \ref{fraction}(c), a
relaxation process right after the critical point
T$_{\text{s},[111]}$ has been observed. Although the fractions of
fcc and hcp phases are reduced, the arrangement of atoms becomes
more inerratic, as shown in Fig. \ref{111}(c) and \ref{111}(g). At
the end of loading, the stacking fault arrangement comes out in the
interior of thick samdwich structure. Compared to the [011] loading,
both fcc and hcp nucleation crystalline planes in [111] loading are
(101)$_{\text{bcc}}$ and (011)$_{\text{bcc}}$. Besides, the grain
boundary planes ($\bar{1}$10)$_{\text{bcc}}$ in [111] loading are
parallel, instead of perpendicular (in the case of [011] loading),
to the compression orientation.

On the whole, for [001] loading, the
$\alpha\mathtt{\rightarrow}\epsilon$ transition happens in the whole
body of the sample and fcc structure only appears as the
intermediate structure. However, for [011] and [111] loadings, the
fcc and hcp structures nucleate simultaneously. Note that for [011]
loading the nucleation is localized. The hcp structure reaches its
maximum at the evolved strain of $\mathtt{\sim}$0.09 and
$\mathtt{\sim}$0.1 for [011] and [111] loadings, respectively.
Further increase of strain will result in the decease (increase) of
hcp (fcc) weight. A visible difference between [011] and [111]
loadings is that for the latter, there occurs an evident relaxation
process after the critical point.

\subsection{Radial distribution function analysis}

\begin{figure}[tbp]
\begin{center}
\includegraphics[width=0.8\linewidth]{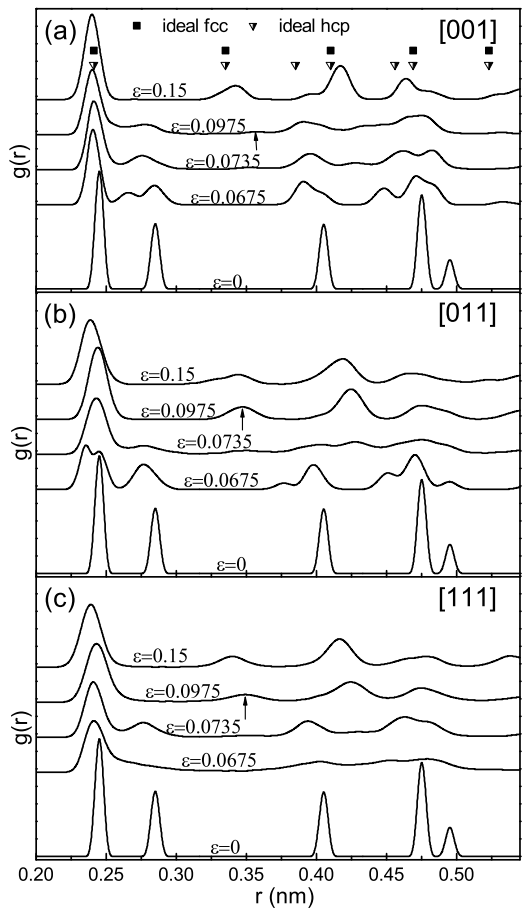}
\end{center}
\caption{Radial distribution functions (RDFs) under different
strains $\varepsilon$ for crystals compressed along [001] (a), [011]
(b), and [111] (c) directions. As a reference the ideal positions of
the fcc and hcp structures, as well as the distribution function for
the unstrained bcc ($\varepsilon$=0), are shown.} \label{rdf}
\end{figure}

The slip investigation of atoms must presume upon radial
distribution function (RDF) analysis and phase mass fraction
analysis. Figure \ref{rdf} displays the RDFs under different
compressions, where the strains of all three samples are from 0 to
0.15. The ideal positions of fcc and hcp structures have also been
plotted. Along with the increase of strain the first peak in all
three samples deviates towards left gradually, while the second peak
deviates towards the first peak and disappears finally. After the
strain becomes bigger than 0.0975, along with the disappearance of
the second peak, a new peak appears, with the position (shown by the
arrows in Fig. \ref{rdf}) exactly the same as that of the second
peak of the fcc/hcp phases. Therefore, these changes in RDFs
definitely signify atomic transformation to the close packed phases.
Moreover, the evident broadening of the peaks in [011] and [111]
loadings after transition is caused by the temperature effect. The
third and fourth peaks at $\varepsilon$=0.15 in Fig. \ref{rdf}(a)
obviously trend to split; this is prominently different from that in
Figs. \ref{rdf}(b) and \ref{rdf}(c). This further illustrates that
sample A is mainly featured by the hcp phase at the end of loading.
All these analyses indicate that the phase distribution and the
mechanism of local slip in three samples are different in the
details of transition process.

\section{CONCLUSIONS}

In summary, we have simulated the process of
$\alpha\mathtt{\rightarrow}\epsilon/\gamma$ structural phase
transition in bcc single crystal iron under uniaxial compression
along three typical bcc orientations by MD method. It is found that
samples B and C experience a more evident hardening process compared
with sample A before phase transition. The temperatures for [011]
and [111] loadings increase more highly than that for [001] loading.
The mixed phase interval of sample B is much shorter than other two
samples, which is caused by the fact that the compression
orientation belongs to the nucleation planes \{110\}$_{\text{bcc}}$.
The transition pressure is the same for all three samples with a
value of $\mathtt{\sim}$14 GPa, over which the nucleation of hcp and
fcc atoms appears and grows into different morphologies. For [001]
loading, the $\alpha\mathtt{\rightarrow}\epsilon$ transition happens
in the whole body of the sample and hcp atoms nucleate on
(011)$_{\text{bcc}}$ planes to form laminar structure; the fcc phase
only appears intermediately. For [011] and [111] loadings, however,
(i) nucleation of fcc and hcp structures occurs simultaneously; (ii)
the hcp mass fraction reaches its maximum at the evolved strain of
$\mathtt{\sim}$0.09 and $\mathtt{\sim}$0.1 and further increase of
strain will result in the decease (increase) of hcp (fcc) weight;
(iii) the fcc and hcp atoms form stacking fault structure and
relatively thick samdwich structure at the end of loading. Note that
for [011] loading the nucleation is localized. The phase
distribution and the mechanism of local slip in three samples are
different in detail.

\begin{acknowledgments}
This study has been supported by the Foundations for Development of
Science and Technology of China Academy of Engineering Physics under
Grants No. 2008B0101008.
\end{acknowledgments}

\end{document}